\documentclass[11pt,a4paper]{article}
 \usepackage[dvips]{graphics}
 \usepackage{graphicx}
 \usepackage{afterpage}
 \usepackage{enumerate}
  \usepackage{longtable}

\newcount\longrefs
\def\aap{\ifnum\longrefs=1 {Astron.\ Astrophys.}\else
{A\hbox{\rm \&}A}\fi}
\def\aapr{\ifnum\longrefs=1 {Astron.\ Astrophys.\ Rev.}\else
{A\hbox{\rm \&}AR}\fi}
\def\aaps{\ifnum\longrefs=1 {Astron.\ Astrophys.\ Suppl.}\else
{A\hbox{\rm \&}A Suppl.}\fi}
\def\aj{\ifnum\longrefs=1 {Astron.\ J.}\else
{AJ}\fi}
\def\ao{\ifnum\longrefs=1 {Applied Optics}\else
{Appl.\ Opt.}\fi}
\def\aspcs{\ifnum\longrefs=1 {Astron.\ Soc.\ Pacific Conf. Series}\else
{ASP Conf.\ Ser.}\fi}
\def\apj{\ifnum\longrefs=1 {Astrophys.\ J.}\else
{ApJ}\fi}
\def\apjl{\ifnum\longrefs=1 {Astrophys.\ J. Lett.}\else
{ApJ}\fi}
\def\aplett{\ifnum\longrefs=1 {Astrophys.\ J. Lett.}\else
{ApJ}\fi}
\def\apjs{\ifnum\longrefs=1 {Astrophys.\ J. Suppl.}\else
{ApJS}\fi}
\def\apss{\ifnum\longrefs=1 {Astrophys.\ and Space Science}\else
{Astrophys.\ Space Sci.}\fi}
\def\araa{\ifnum\longrefs=1 {Ann.\ Rev.\ Astron.\ Astrophys.}\else
{ARA\hbox{\rm \&}A}\fi}
\def\azh{\ifnum\longrefs=1 {Astronomicheskii Zhurnal}\else
{Astron.\ Zhur.}\fi}
\def\baas{\ifnum\longrefs=1 {Bull.\ Am.\ Astron.\ Soc.}\else
{BAAS}\fi}
\def\bain{\ifnum\longrefs=1 {Bull.\ Astronom.\ Institutes Netherlands}\else
{Bull.\ Astr.\ Inst.\ Neth.}\fi}
\def\gca{\ifnum\longrefs=1 {Geochim.\ Cosmochim.\ Acta}\else
{Geochim.\ Cosmochim.\ Acta}\fi}
\def\grl{\ifnum\longrefs=1 {Geophys.\ Res.\ Lett.}\else
{Geoph.\ Res.\ Lett.}\fi}
\def\iaucirc{\ifnum\longrefs=1 {IAU Circulars}\else
{IAU Circ.}\fi}
\def\ip{\ifnum\longrefs=1 {in press}\else
{in press}\fi}
\def\jgr{\ifnum\longrefs=1 {J.\ Geophys.\ Res.}\else
{J.\ Geophys.\ Res.}\fi}
\def\jrasc{\ifnum\longrefs=1 {J.\ Royal Astron.\ Soc.\ Canada}\else
{JRAS Can.}\fi}
\def\mnras{\ifnum\longrefs=1 {Mon.\ Not.\ Roy.\ Astron.\ Soc.}\else
{MNRAS}\fi}
\def\nat{\ifnum\longrefs=1 {Nature}\else
{Nat}\fi}
\def\pasj{\ifnum\longrefs=1 {Pub.\ Astron.\ Soc.\ Japan}\else
{PASJ}\fi}
\def\pasp{\ifnum\longrefs=1 {Pub.\ Astron.\ Soc.\ Pacific}\else
{PASP}\fi}
\def\physscr{\ifnum\longrefs=1 {Physica Scripta}\else
{Phys.\ Scrip.}\fi}
\def\planss{\ifnum\longrefs=1 {Planetary \& Space Science}\else
{Plan. \& Space Sci.}\fi}
\def\procspie{\ifnum\longrefs=1 {Proc.\ SPIE}\else
{Proc.\ SPIE}\fi}
\def\qjras{\ifnum\longrefs=1 {Quarterly J.\ Royal Astron.\ Soc.}\else
{QJRAS}\fi}
\def\sa{\ifnum\longrefs=1 {Soviet Astron..}\else
{Sov.\ Astron.}\fi}
\def\skytel{\ifnum\longrefs=1 {Sky \& Telescope}\else
{Sky \& Tel.}\fi}
\def\solphys{\ifnum\longrefs=1 {Solar Phys.}\else
{Sol.\ Phys.}\fi}
\def\ssr{\ifnum\longrefs=1 {Space Science Rev.}\else
{Space\ Sci.\ Rev.}\fi}

\begin{document}

 \title{\bf Asymmetry of Lines in the Spectra
of the Sun and Solar-Type Stars
}
\author{\bf V. A. Sheminova}
 \date{}

 \maketitle
 \thanks{}
\begin{center}
{Main Astronomical Observatory, National Academy of Sciences of
Ukraine,
\\Akademika  Zabolotnoho 27,  Kyiv,  03143 Ukraine\\ e-mail: shem@mao.kiev.ua}
\end{center}

\begin{abstract}
{We have analysed the asymmetry of lines Fe I and Fe II in spectra of a solar flux using three FTS atlases and  the HARPS atlas and also in spectra of 13 stars using observation data on the HARPS spectrograph. To reduce observation noise individual line bisectors of each star have been averaged.  The obtained average bisectors in the stellar spectra are more or less similar to the shape C well known to the Sun.  In stars with rotation speeds greater than 5 km/s the shape of the bisectors is more like /. The curvature and span of the bisectors increase with the temperature of the star.  Our results confirm the known facts about strong influence of rotation velocity on the span and shape of bisectors. The average convective speed was determined based on the span of the average bisector, which shows the largest difference between the velocity of cold falling and hot rising convective flows of the matter. It's equal to $-420$ m/s for the Sun as a star. In solar-type stars, it grows from $-150$ to $-700$ m/s with an effective temperature of 4800 to 6200 K, respectively. For stars with greater surface gravity and greater metallicity, the average convective velocity decreases. It also decreases with star age and correlates with the speed of micro and macroturbulent movements.  The results of solar flux analysis showed that absolute wavelength scales in the FTS atlases used coincide with an accuracy of about $-10$ m/s, except for the atlas of Hinkle, etc., whose scale is shifted and depends on the wavelength. In the range from 450 to 650 nm, the scale shift of this atlas varies from $-100$ to $-330$~m/s, respectively, and it equals on average of $-240$ m/s. The resulting average star bisectors contain information about the fields of convective velocities and may be useful for hydrodynamic modeling of stellar atmospheres in order to study the characteristic features of surface convection.}
\end{abstract}


\section{Introduction }

In stellar spectra asymmetry of absorption lines  may arise due to  dark spots, bright points, and bright plages, as well as due to oscillations, pulsations, granulation, and line blending. The main cause of asymmetry in solar-type stars is granulation \cite{DRAVINS1981}.
It is directly observed in images of the solar surface as a granulation pattern.
For main-sequence stars direct
observation of granulation is impossible, however, signatures of stellar granulation can be studied by analyzing
the asymmetries of the absorption line profiles. By studying the asymmetry of many lines, we can obtain
information about the characteristics of surface convection in stars. The observation of spectral
lines in stars, as compared with the Sun, has a number of limitations.
Firstly, in cases of stars we observe a disk-integral
flux, which leads to the weakening of the Doppler shifts. Secondly, the absolute line shifts cannot be measured
due to the lack of data on the exact radial velocities of many stars. Thirdly, the rotation of the stars
affects the shape of the line profiles and alters the asymmetry. Fourthly, spectral lines become weaker in
low-resolution stellar spectra, and profile distortions increase. This should be borne in mind when studying
the asymmetry of lines in stars.

The appearance of granulation on the stellar surface is caused by convection in the surface layers,
which creates wide upward flows of hot matter, so-called granules, and cold downward flows, or intergranules.
Granules make a blue contribution to the spectral line due to the matter's ascent velocity. Cold
intergranules are narrow because of the smaller area and they make a red contribution due to the matter's
descent velocity. The combined effect of all contributions integrated over the stellar surface creates asymmetric
line profiles  \cite{DRAVINS1981}.  
Photospheric velocities that are directly caused by convection and affect the asymmetry of lines are called
convective or granulation velocities. According to [22], the average velocity of
the distribution of granulation velocities is the average ascent velocity of granules $V=V_{\rm col}- V_{\rm hot}$,
where $V_{\rm col}$  and $V_{\rm hot}$  velocity of granules  are the velocities of cold and hot matter, respectively;
the dispersion of this distribution
is the so-called macroturbulent velocity $V_{\rm mac}$. Simple modeling showed that the average granulation
velocity affects the asymmetry, while the velocity dispersion affects the line width, and always $V < V_{\rm mac}$.
As noted in \cite{1982ApJ...255..200G},  
asymmetry is controlled by three parameters: (1) average velocity of granules and intergranules, (2) brightness
contrast between granules and intergranules, and (3) the fraction of the
stellar disk area occupied by granules in relation to intergranules. For a star, (2) and (3) are combined
into a flux contrast $(F_{\rm hot}-F_{\rm col})/(F_{\rm hot}+F_{\rm col})$, where $F_{\rm hot}$ is the flux
emitted by all granules, and $F_{\rm col}$ is the
flux emitted by all intergranules on the stellar disk. The higher the degree of correlation between the velocity
and temperature in the granules and in the intergranules, the stronger the asymmetry of the observed
line. The rotation velocity $v \sin i$ also affects the asymmetry when it is higher than $V_{\rm mac}$. According to
\cite{2008A&A...492..841R},   
the asymmetry is to a lesser extent influenced by the abundance of chemical elements and the activity of
stars, and the latter decreases the line asymmetry in comparison with inactive stars.

The asymmetry of the profiles is quantitatively characterized by the bisector, which represents the
dependence between convective shifts and the relative flux of radiation at each point of the line profile.
The most complete picture of convective shifts is seen for strong lines. This is due to a significant difference
in physical conditions at different effective formation depths of each segment of the line where the
bisector is measured. Early studies of the solar spectrum
\cite{1984SoPh...93..219B, 
        DRAVINS1981,
     1977SoPh...53..353K, 
    1983IAUS..102..149L}  
established that the bisector has a
typical shape similar to the letter C. The upper part of the profile near the continuum has a small blue
shift due to significant contributions of radiation from intergranules in the deep layers of the photosphere.
The largest blue shift occurs closer to the middle of the profile due to the predominant fraction of radiation
from hot rising granules. The core of a strong line has a smaller blue shift because the radiation contrast
between granules and intergranules decreases in the upper photospheric layers of the core formation. The
asymmetry of the line profile varies depending on the line depth and atomic parameters, such as chemical
composition, excitation and ionization potential, wavelength, oscillator strength, i.e., the parameters that
determine the effective depth of line formation in the atmosphere.

Stellar bisectors show significant differences both in the shape and the magnitude of the blue span with
variations in effective temperature, luminosity class, or gravity. According to the results of stellar spectra studies
\cite{1999ApJ...526..991A,   
         1987A&A...172..211D,   
        1999ASPC..185..268D,   
        2008A&A...492..199D,    
        1982ApJ...255..200G,  
       2005PASP..117..711G,  
        1989ApJ...341..421G,  
       1985PASP...97..543G,  
        2008A&A...492..841R},   
the bisector does not show the C-shape in all cold FGK stars of the
main sequence; this shape is often slightly distorted, and sometimes only the upper part of the letter C
is visible. The differences in shape from one star to another depend on the strength of granulation and the
structure of the atmosphere. The span or shift of the bluest point of the bisector, which characterizes the
average granulation velocity of the star,  decreases from F- to K-stars. For giants, the span is larger than
for dwarfs. The brighter the star (or the greater the luminosity, or the smaller the gravity), the lower the height
of the bluest point of the bisector, and the higher the granulation penetrates into the atmosphere. The spectra
of stars with a very low metal abundance reveal a significantly greater line asymmetry and a wider range of
velocities than stars of the same class with solar metallicity
\cite{1999ApJ...526..991A}.   
This fact is interpreted as a signature of a low metal
abundance and, therefore, a low opacity in convective atmospheres. There is also another feature of asymmetry:
if we proceed along the main sequence towards higher temperatures, the line bisectors in the F0 zone
change their tilt to the opposite direction. This boundary separating the two modes is called the granulation
boundary
 \cite{ 1989ApJ...341..421G, 1986PASP...98..499G}.   
It is assumed that convection in hot F0V stars is qualitatively different: granules are small
and quickly rise upward, while intergranules are large and descend slowly.

To better understand stellar granulation and its effect on the observed spectra, hydrodynamic modeling
of the solar and stellar atmospheres is performed, which takes into account atmospheric inhomogeneities
associated with the granulation phenomenon. The first studies on granulation modeling
\cite{2000A&A...359..729A,  
1994A&A...291..635A, 
1990A&A...228..203D}  
predicted that the energy flux passing through the atmosphere is the most important parameter that affects
the magnitude of blue shifts and, therefore, the span of the bisector. The correlation between temperature,
velocity, and density is weaker in cold dwarfs than in hotter stars. Subsequent studies on modeling the convective
envelopes of stars, e.g.,
 \cite{2002ApJ...567..544A,  
      2013A&A...550A.103A, 
      2013A&A...558A..49B, 
      2009A&A...501.1087R},  
confirmed and explained many observational features of
bisectors. Despite the great progress in modeling, theoretical calculations were carried out only for some
FGK stars using a limited set of spectral characteristics and were verified against a limited set of observations.
It is necessary to study a larger number of stars with hotter and colder temperatures. Careful validation
of the models on high-quality observed spectra is required. It is especially important to verify the
models for coincidence with the bisectors of many observed lines. This should be done so that the granulation
properties predicted by the models can be guaranteed to be fully reliable. Therefore, obtaining new
data on the asymmetry of the observed lines and on their bisectors for stars with different characteristics
remains an important problem in the physics of stellar atmospheres. Solving this problem requires high
quality stellar spectra and careful selection of absorption lines to reduce the bisector errors.

The aim of this study is to measure the line bisectors in the spectra of solar-type stars with different
characteristics and to trace the change in their shape with the variations in the stars' parameters. In other
words, this study is dedicated to the second signature of stellar granulation, i.e., line asymmetry. The first signature
of granulation is the broadening of the line profiles by macroturbulent motions, and the third signature of granulation
is the blue shift of lines due to the dominant contributions of hot granules. The well-known parameter
of classical macroturbulence shows the strength of granulation in one or another star. In our previous
study
\cite{2019KPCB...35..129S}, 
we already investigated the first signature of granulation in the stars that are analyzed here. This
enables us to perform a comparison of convective and macroturbulent velocities and confirm the existing
understanding of granulation processes and its effect on spectral lines.

\section{Initial data and methods}

{\bf Observation data.} Table 1 shows the basic data for the stars whose spectra we will analyze. These stars
are from the Calan-Hertfordshire Extrasolar Planet Search (CHEPS) sample. The main characteristics of
the stars, such as effective temperature $T_{\rm eff} $, surface gravity $\log g$, and metallicity [M/H], were taken from
 \cite{2017MNRAS.468.4151I}, 
 the data on mass $m$ and age from
\cite{2019A&A...621A.112P}, 
and the data on macroturbulent velocity $V_{\rm mac}$   and rotational
velocity $v \sin i$ from
\cite{2019KPCB...35..129S}. 
The spectral types correspond to the data from the SIMBAD database
(http://cdsportal.u-strasbg.fr). All the stars are single and inactive
($\log R_{\rm HK} \leq -4.5$ dex). The $T_{\rm eff}$  range is
small, from 4800 to 6200 K. Some of the stars are slightly enriched in metals with [M/H] from 0.06 to 0.34,
while others have a small metal deficit from $-0.05$ to $-0.15$.

\begin{table}[!hb]
\centering
 \caption{\small Values of the parameters of the stars ($n$ is the number of lines; $V$ is the average convective velocity in the Mg scale)}
 \vspace {0.3 cm}
\footnotesize
\begin{tabular}{llccccccccc}
\hline\hline
HD&Type & $T_{\rm eff}$&$\log g$&[M/H]&$M/M_\odot$&Age   & $V_{\rm mac}$  & $v \sin i$  & $n$ & $V$\\
  & &         (K)   &        &     &           &(Ga)& (km/s) & (km/s)&  &m/s\\
\hline
189627& F7 V      &6210   &4.40   &~~0.07   &1.244& 4.0 &5.52 &5.93& 31 &  $-$606\\
92003 & F0 V      &6158   &5.10   &~~0.27   &1.087& 7.2 &2.20 &2.09& 37 &  $-$573\\
93849 & G0/1 V    &6153   &4.21   &~~0.08   &1.268& 3.5 &2.92 &3.05& 41 &  $-$696\\
158469& F8/G2 V   &6105   &4.19   &$-$0.14  &1.223& 2.0 &3.61 &3.10& 44 &  $-$686\\
127423& G0 V      &6020   &4.26   &$-$0.09  &1.107& 3.1 &2.90 &2.53& 34 &  $-$442\\
6790  & G0 V      &6012   &4.40   &$-$0.06  &1.089& 3.5 &3.16 &2.94& 32 &  $-$511\\
102196& G2 V      &6012   &3.90   &$-$0.05  &1.395& 3.0 &4.26 &3.56& 36 &  $-$630\\
102361& F8 V      &5978   &4.12   &$-$0.15  &1.250& 2.0 &5.62 &5.03& 47 &  $-$600\\
147873& G1 V      &5972   &3.90   &$-$0.09  &1.493& 2.6 &5.95 &6.51& 34 &  $-$586\\
Sun~~~& G2 V      &5777   &4.44   &~~0.00   &1.000& 4.6 &2.11 &1.84& 61 &  $-$456\\
38459 & K1 IV-V   &5233   &4.43   &~~0.06   &0.882& 9.0 &3.20 &1.85& 37 &  $-$220\\
42936 & K0 IV-V     &5126   &4.44   &~~0.19   &0.881&12.0 &1.74 &0.97& 26 &$-$338\\
221575& K2 V      &5037   &4.49   &$-$0.11  &0.823& 6.0 &2.79 &1.89& 36 &  $-$152\\
128356& K2.5 IV   &4875   &4.58   &~~0.34   &0.824&15.5 &1.74 &1.01& 19 &  $-$220\\
\hline
\end{tabular}
\end{table}
\noindent

The spectra of the stars were obtained with the HARPS (High-Accuracy Radial Velocity Planetary
Searcher) spectrograph at La Silla in Chile
 \cite{2009MNRAS.398..911J}. 
 The signal-to-noise ratio is above 100. The spectral resolution
$R\approx 120000$  is rather low for this problem; however, according to
\cite{2005PASP..117..711G}, 
this is sufficient, since the
nominal spectral resolution should be at least 100000.

{\bf List of spectral lines.} The lines were selected visually using a graphic solar atlas
\cite{2011ApJS..195....6W}  
with a high resolution
$R \approx 700 000$. To reduce uncertainties due to noise, bisectors were measured in the
$F_{\lambda}{/}F_{\rm c}$ range from 0.1 to 0.95. Here, $F_{\lambda}$ and $F_{\rm c}$ are
the radiation fluxes in the line and the continuum, respectively. The
selected lines were in the wavelength range from 450 to 650 nm. The number of lines varied from star to
star, because new blends appeared in the line profiles with decreasing $T_{\rm eff}$. The presence of subtle blends
in the profiles was checked by analyzing the bisectors and comparing them with the mean bisector. As a
result of this selection, the final lists of lines in different stars mainly differed in the range of very weak and
strong lines. The number of lines $n$ is given in Table 1.

{\bf Measurement of the central wavelength of a line.} To accurately determine the wavelength of a line in the
spectrum of a star is only possible for a symmetrical profile. In reality, stellar spectral lines are always
asymmetric and distorted by blends and noise, which makes it impossible to accurately and unambiguously
calculate their central wavelength. There are various procedures for measuring the line wavelength
\cite{1999PASP..111.1132H}. 
In most cases, the wavelength of the line is determined by fitting polynomials
of the 2--4 order by the number of points nearest to the minimum. We have chosen, in our opinion,
a more error-tolerant method using the center of gravity of the line core
 \cite{1997KPCB...13e..65B}.  
The accuracy of this
method is not affected by a strong or weak, wide or narrow line in the core. Mainly what matters is the
interpolation step of the profile. In our case, it was equal to 0.1 pm.

{\bf Calculation of the convective line shifts.} For stars, it is difficult to determine convective line shifts because
the true radial velocity of the stellar center of mass is not known accurately enough. To exclude the influence
of the radial velocity, the zero point of the shifts is established using absolute solar shifts, as was done in
\cite{2018ApJ...857..139G}, 
or an arbitrary zero point is adopted using the shifts of the core wavelengths of the strongest lines. In
\cite{1983IAUS..102..149L},  
to determine the convective line shifts observed in magnetic formations on the solar surface, the strong line
Mg I  ${\rm b}_2$ 517.27 nm was used (hereinafter referred to as the ``Mg line'') based on the assumption that the core
of this line forms high in the atmosphere, where the brightness contrast nearly disappears and its convective
shift will be almost zero. This line was also used in
\cite{1997KPCB...13e..65B} 
a long with a group of strong iron lines to determine
the zero point of the shifts of the observed lines in magnetic formations. In the study of K-dwarfs
\cite{2008A&A...492..841R},  
the shifts of the cores of the strongest lines of iron were also used on the basis of the assumption that they are
equal to zero, which was based on the results obtained for the solar spectrum in
 \cite{1998A&AS..129...41A}. 

In this study, we used the Mg line. This allowed us to eliminate Doppler shifts due to the radial velocity
of the star relative to the observer on the Earth, gravitational displacement, and other displacements if they
were not taken into account in the calibration of the observations as well as errors in the method of measuring
the central wavelength. Such a zero point of the shift scale will contain an error that is equal to the
intrinsic convective shift of the Mg line. Such a scale for stellar spectra will henceforth be called the Mg
scale. In this scale, the convective shift of the spectral line is easily calculated relative to the laboratory
wavelength in units of velocity using the following formula
\begin{equation}
  V = c(\lambda_{\rm obs} - \lambda_{\rm lab}){/}\lambda_{\rm lab} -
     c(\lambda_{\rm obs}^{\rm Mg} - \lambda_{\rm lab}^{\rm Mg}){/}\lambda_{\rm lab}^{\rm Mg}.
\end{equation}
Here $\lambda_{\rm obs}$  is the measured central wavelength of the observed line in the spectrum of the star,
$\lambda_{\rm lab}$ is the
laboratory wavelength. The superscript Mg indicates the Mg line. We have applied this formula to all stars
and tested it using the Sun's spectrum.

{\bf Laboratory wavelengths.} To measure convective shifts, we used the results of laboratory studies of wavelengths
\cite{1994ApJS...94..221N}.  
The catalog of laboratory wavelengths presented in \cite{1994ApJS...94..221N} is currently one of the most accurate
and contains a complete list of iron lines. All the line wavelengths are divided into four categories depending
on their errors (0.04 to 1 pm and larger). Most lines have an error below 0.05~pm, and approximately half of
them are below 0.1 pm (or 60 m/s for a wavelength of 500.0 nm). For weaker lines ($ F_{\lambda} {/} F_{\rm c} >  0.7 $), the error of laboratory wavelengths is approximately 2.5 times larger than for stronger lines.

{\bf Plotting of the bisectors.} To obtain a bisector, we first interpolate the line profile using a cubic spline
with a step of 0.01 pm. We then calculate the midpoints of the horizontal lines drawn from the measured
point on the blue side of the line profile to the interpolated point on the red side of the profile. By connecting
the resulting series of midpoints, we obtain a curve that is the bisector of the line. The ordinate
axis on the bisector plots is the radiation flux $F_{\lambda} {/} F_{\rm c}$ normalized to the continuum,
and the abscissa axis is the wavelength shifts of the measured profile points calculated according to
the above formula and converted to velocity units.

Since more and more blends appear in the spectra when proceeding to cooler stars, the line profiles
become more noisy, their bisectors are more distorted, and the number of blendless lines decreases. For
this reason, reliable results cannot be obtained by analyzing individual bisectors in the spectra of many
stars. There is a practice to remove any noise by averaging the bisectors. For example, it was shown
 \cite{1999ApJ...526..991A, 2008A&A...492..199D} 
that it is possible by averaging bisectors to obtain a statistically reliable result by discarding points that
deviate sharply from the mean bisector. We applied this method as well. All abrupt outliers were considered
erroneous and removed from the averaging or the top and/or bottom parts of the bisector were cut off
if they contained a large outlier. In such cases, there is inevitably some degree of subjective judgment, so
we acted with caution. The mean bisector was calculated as the arithmetic mean of the positions of those
bisectors that contribute at each absorption depth with a step of 1\% of the intensity of the continuum. For
each star, we averaged the individual bisectors in the Mg scale and then smoothed the mean bisector.

\section{Line bisectors in the solar flux spectrum}

{\bf Individual and mean bisectors.} We performed an analysis of solar line bisectors to add the Sun to our
sample of solar-type stars to ensure the validity of our results. For this purpose, we used the solar atlas
\cite{2013A&A...560A..61M}, 
that represents the spectrum of the Moon-reflected solar flux obtained at HARPS in 2010 with a resolution
$R \approx 120 000$  in the spectral range from 476 to 585 nm and with a small gap in the range of 530--534 nm. We
will refer to this atlas as HARPS. This atlas was calibrated using an ideal laser frequency comb (LFC) calibrator.
The authors of this atlas claim that the wavelength solution is the most accurate of all currently
possible and requires no other correction. We will use the HARPS atlas for the comparison with the results
of the analysis of stellar spectra obtained with the same spectral resolution.

In addition, three more high-resolution solar flux atlases obtained with FTS (Fourier-Transform
Spectrometers) were used. The HINKL FTS atlas
\cite{2005ASPC..336..321H} 
provides convenient access to the KURUCZ atlas \cite{1984sfat.book.....K}
with a resolution $R \approx 340500$, 521360, and 522900 for the wavelength ranges of
401.9--473.8, 473.8--576.5, 576.5--753.9 nm, respectively. The NECKL FTS atlas
 \cite{1999SoPh..184..421N} 
provides a high-quality spectrum with
$ R \approx 400 000 $  for the center of the disk and averaged spectrum over the entire disk. The IAG FTS atlas
 \cite{2016A&A...587A..65R} 
has a very high resolution $R \approx 1000 000$  and an accurate wavelength scale. Its systematic deviation from the
HARPS scale is approximately $-5 \pm 5$  m/s and it has no significant trend in the range of 480--580 nm. We
will further use the IAG atlas as a reference for the comparison with the other atlases of the solar spectrum.

 \begin{figure}[!ht]
 \centerline{
 \includegraphics   [scale=1.1]{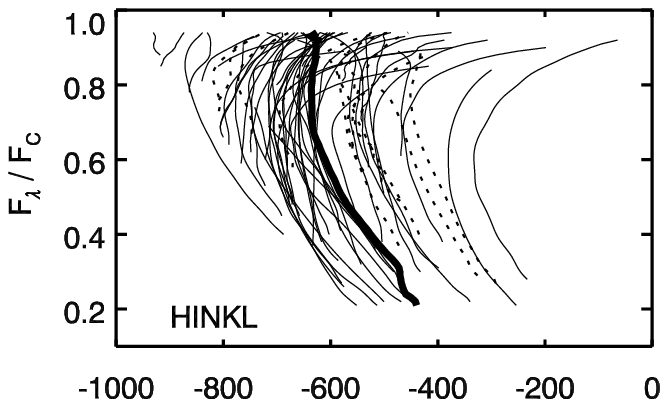}
 \includegraphics   [scale=1.1]{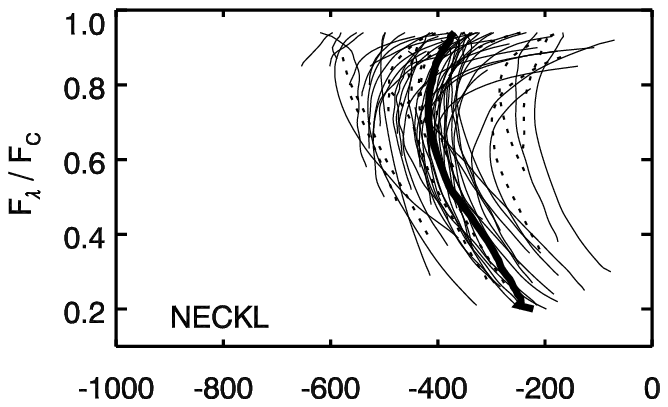}
 }
 \centerline{
 \includegraphics   [scale=1.1]{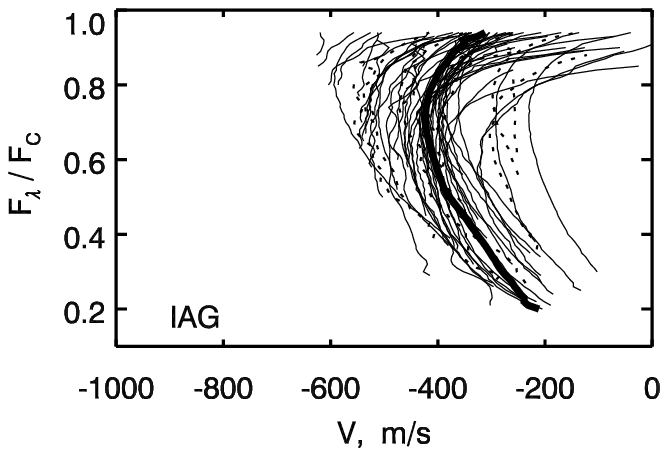}
  \includegraphics  [scale=1.1]{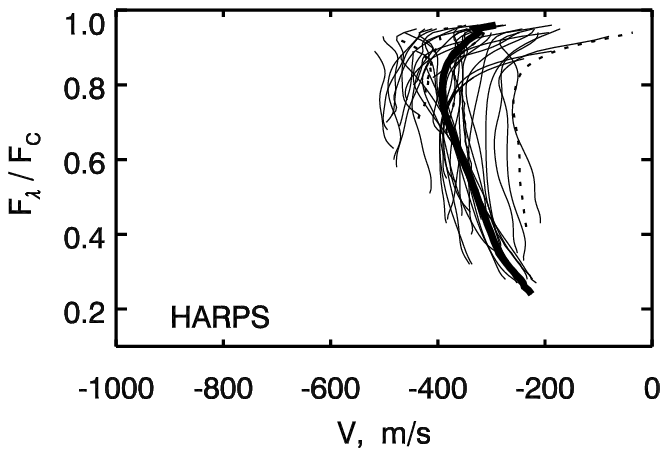}
  }
  \caption {\small
Individual and mean bisectors of iron lines measured in the absolute wavelength scale of four spectral atlases of the
solar flux HINKL, NECKL, IAG, and HARPS: the thin lines are the bisectors of the Fe I lines, the dashed lines are the
bisectors of the Fe II lines, the bold lines are the mean bisectors.
}
 \end{figure}
\begin{figure}[h!b]
 \centerline{
 \includegraphics   [scale=01.1]{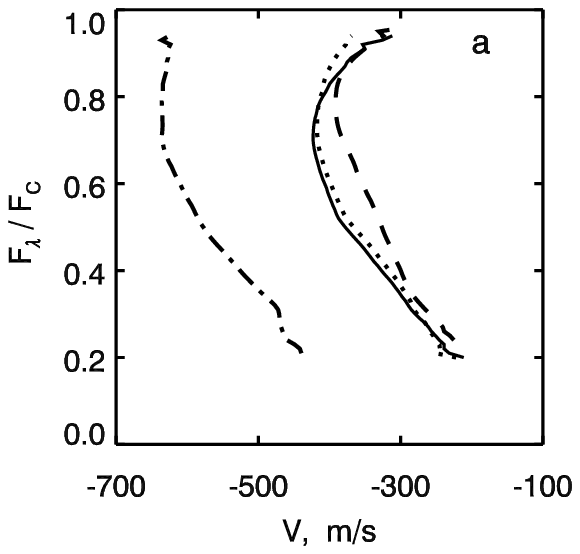}
 \includegraphics   [scale=01.1]{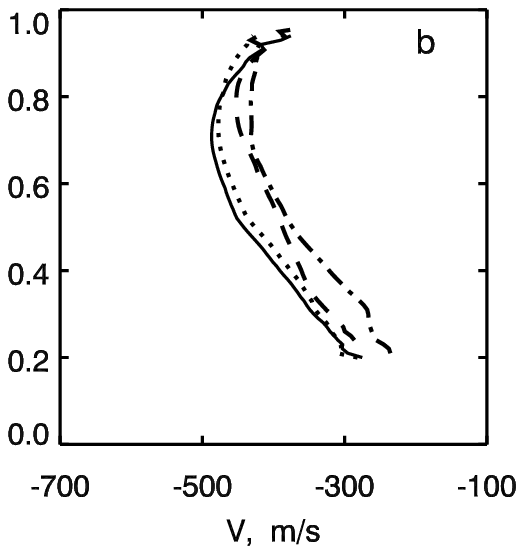}
 }
   \caption {\small
Mean bisectors of the lines in the solar flux spectrum (a) in the absolute scale and (b) in the Mg-scale according
to the observations from four atlases: IAG (solid line), HINKL (dash-dotted line), NECKL (dotted line), and HARPS
(dashed line).
}
 \end{figure}

We have measured bisectors for 61 lines. All the bisectors of these lines, as well as the mean bisector,
are shown in Fig.~1 in the absolute scales of each atlas. As can be seen, the range of blue shifts is the same
for the IAG and NECKL atlases, while that for HARPS is narrower, and it is significantly broadened and
shifted towards the blue part of the spectrum for the HINKL atlas. Figure~2a shows the mean bisectors of
the four atlases in absolute scales. The average convective velocity was determined from the span of the
mean bisector. For NECKL and IAG, it is $-418$ and $-423$ m/s, respectively. For the HARPS atlas, it is
slightly lower ($-395$ m/s) due to the lower spectral resolution. The lower-resolution solar spectrum weakened
by the reflection from the Moon shows smoother line profiles with less pronounced asymmetry. The
authors of the HARPS spectrum provided the solar radiation flux that was not normalized to the continuum,
so we used the local continuum for each individual line. This may partially affect the shape of the
bisector near the continuum. For the HINKL atlas, the bisector span turned out to be the largest and,
accordingly, the average convective velocity turned out to be large ($-636$ m/s). The shape of this bisector
is also different from the IAG. The bend in the upper part of the bisector is smaller due to increased blue
shifts of the weakest lines with wavelengths greater than 600 nm. Based on the comparison results, we
came to the conclusion that the wavelength scales in the IAG, NECKL, and HARPS atlases nearly coincide
within the analysis errors and are quite accurate. The average convective velocity in the photosphere
of the Sun as a star measured from the data of these atlases is approximately $-420$ m/s. There appears to
be an issue with the wavelength scale in the HINKL atlas, which will be discussed below.

{\bf Errors of the absolute scales of solar atlases.} We compared the absolute shifts of the mean bisectors of
each atlas with the IAG and, thus, estimated the error of the absolute scales in the wavelength range from
450 to 650 nm. The NECKL and HARPS atlases have an insignificant error of approximately $-10$ m/s
(Fig.~2a), while the error in the HINKL atlas is $-240$ m/s. This infers that the absolute scales of the IAG,
NECKL, and HARPS atlases can be considered to be reliable, while the HINKL scale cannot. The analysis
of the shifts of individual bisectors showed that the scale shift in the HINKL atlas relative to IAG
 \begin{figure}[h!t]
 \centerline{
 \includegraphics   [scale=1.2]{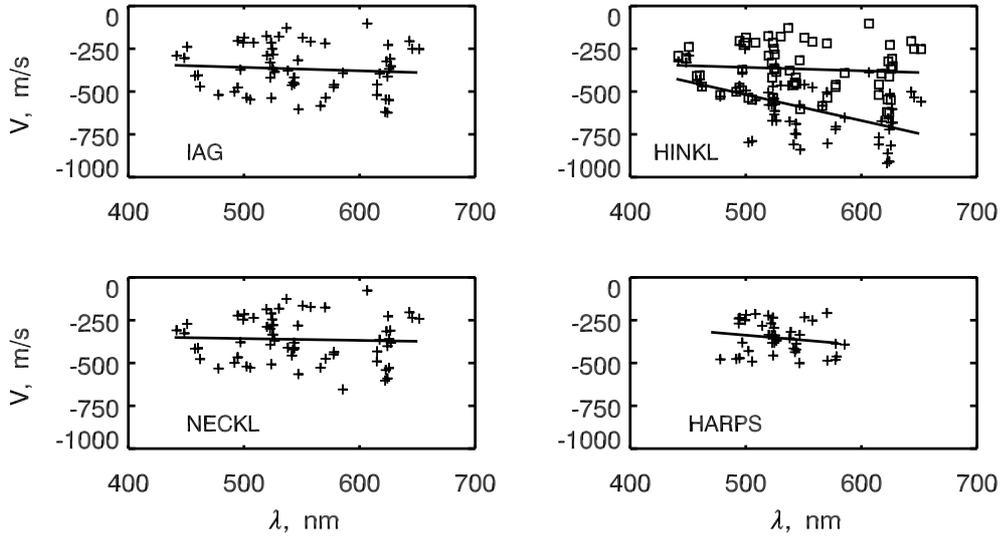}
     }
  \caption {\small
Absolute convective shifts of the iron lines in the solar flux spectrum according to the IAG, HINKL, NECKL,
and HARPS atlases depending on the wavelength (the squares show the line shifts from the IAG atlas for comparison).
}
 \end{figure}
 \begin{figure}[h!b]
 \centerline{
  \includegraphics   [scale=1.]{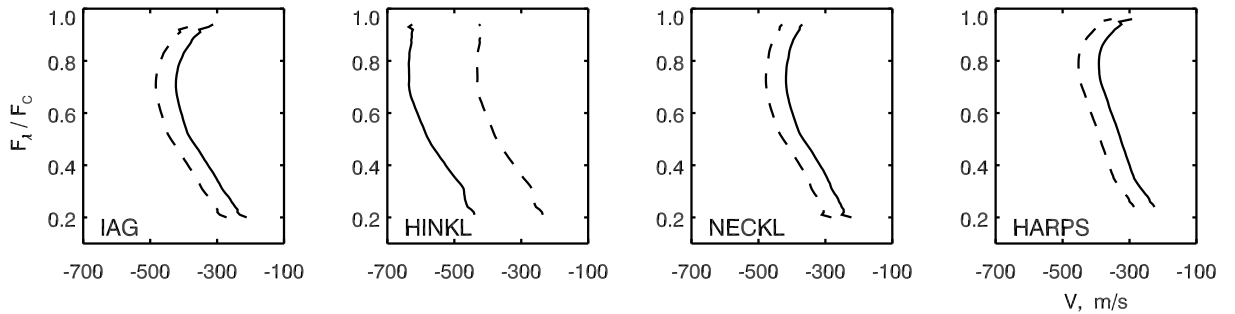}
     }
  \caption {\small
Mean bisectors of the lines according to the data of the IAG, HINKL, NECKL, and HARPS atlases in the absolute
scales (solid lines) and Mg scales (dashed lines).
}
 \end{figure}
depends on the line wavelength. Figure~3 demonstrates this dependence for the HINKL atlas, while the
IAG, NECKL, and HARPS atlases show only an insignificant trend, which is a consequence of this line
selection. In the HINKL atlas, blue line shifts increase with wavelength. The greater the wavelength, the
greater the bisector shift. The difference between the shifts in HINKL and IAG is approximately $-100$, $-220$,
and $-330$ m/s for 450, 550, and 650 nm, respectively. The growth of the blue shifts with wavelength contradicts
the physical sense. Just the opposite should be the case, since opacity increases with wavelength,
the contrast between the granules and intergranules decreases, and the blue shift should decrease. As
shown in [25], there is hardly any dependence of blue shifts on wavelength for lines of different strengths.
However, if we select lines of equal strength, the blue shifts of these lines decrease with wavelength. That
being said, weak lines show less dependence than strong lines. We used lines of different strengths, so there
should be no dependence on the wavelength in our case. This suggests that the wavelength scale of the
spectrum in the HINKL atlas is, apparently, not accurate enough. It was noted in [20] that the zero point
in the HINKL atlas was reset, apparently in order toalign the average positions of the Fe I lines with laboratory
values [34]. According to the data presented [20], the shift between HINKL and IAG in the range
600--630 nm is $-336$ m/s. This agrees with our estimates for that particular range.

{\bf Verification of the Mg scale.} Using the solar data, we can verify the Mg scale plotted with respect to the
Mg line. Since all the solar atlases have absolute scales, we can trace possible uncertainties in the Mg scale.
Figure 4 shows the mean bisectors in this scale (dashed line) and the bisectors in the absolute scale for
each atlas (solid line). The bisectors in the Mg scale are shifted by the corresponding absolute shift of the
Mg line in each atlas, the value of which depends on the calibration accuracy of the atlas. In absolute
scales, this shift is $-204$ m/s (HINKL), 64 m/s (IAG), 60 m/s (NECKL), and 61 m/s (HARPS). The
close values obtained for three atlases allow us to conclude that the Mg line in the solar spectrum has
a small red shift of $ \approx 60$  m/s. When comparing the mean bisectors of four atlases in the Mg scale (Fig. 2b),
the difference between HINKL and IAG is significantly reduced. Almost all the mean bisectors are in satisfactory
agreement within the analysis error. Therefore, we can conclude that the use of the Mg line for
constructing the wavelength scale allows reducing possible errors in the central wavelengths of the lines
but introduces an additional error equal to the intrinsic shift of the Mg line core.

 \begin{figure}[h!b]
 \centerline{
\includegraphics [scale=0.8]{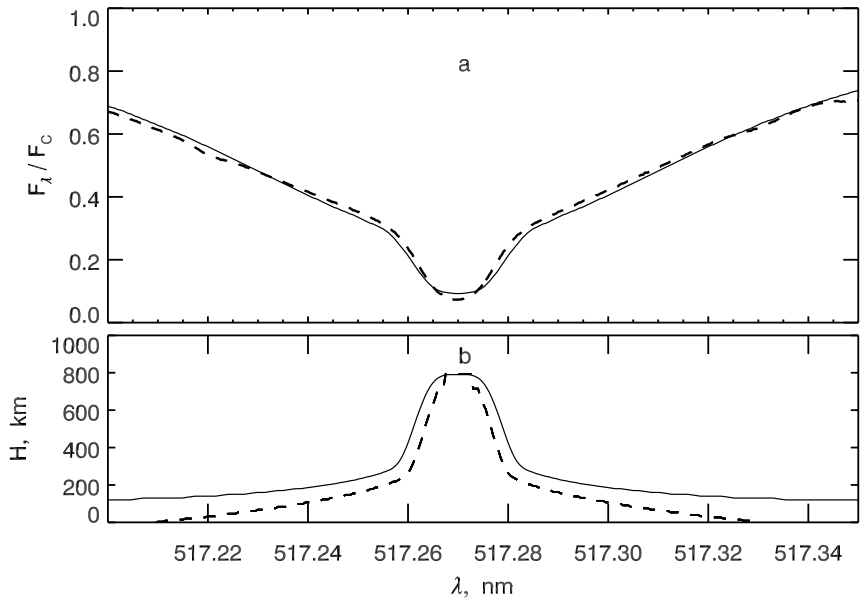}
     }
  \caption {\small
(a) Profile of the Mg line core according to the IAG atlas (dashed line)
and the profile synthesized using the FALC model
 \cite{1993ApJ...406..319F}
(solid line). (b) Profile of effective formation heights calculated from
depression functions (solid line) as well as a profile of heights where
the optical depth in the line is equal to one (dashed line).
}
\end{figure}
It is interesting to note that this intrinsic Mg line shift turned out to be red. In general, the reddening
effect of shifts of very strong lines was previously known both for solar lines
\cite{2002KFNT...18...18S}  
and stellar lines
 \cite{1999ApJ...526..991A}.   
 It is
possible that this small red shift is caused in part by convective currents penetrating into the region
between the upper photosphere and lower chromosphere. Doppler images obtained in the core of this line
show that the granulation structure still exists, but it differs markedly from the granulation pattern seen in
the wings of this line \cite{2011A&A...531A..17R}. According to
\cite{2015KPCB...31...65B, 2004ARep...48..769K}, 
the structure of granulation is preserved up to heights of
approximately 500--700 km, while its characteristics vary with height. The temperature inversion in granules
and intergranules almost always occurs, on average, at heights of 200--300 km, while the inversion of
vertical velocities occurs in most cases at heights $>500$ km. At heights of approximately 700 km, 20\% of
the granulation structures are still observed. These are the heights between the upper photosphere and
lower chromosphere where the core of the Mg line should form. According to \cite{2018MNRAS.481.5675Q}, 
the formation height
of the core of this line calculated for the center of the solar disk is approximately 600 km. The maximum
sensitivity to temperature variations is manifested at a height of approximately 550 km, and that to vertical
velocity variations is at approximately 800~km. We calculated the formation height of the Mg line core in
the solar flux spectrum with the aid of the SPANSAT software
\cite{1988ITF....87P...3G} 
using the depression contribution
functions
\cite{2015arXiv150500975G}. 
Figure 5 shows the observed and calculated profiles of the Mg line core and the profiles of
effective heights. As can be seen, in the range $F_{\lambda}{/} F_{\rm c} = 0.1$--0.2 ,
the core forms at heights of 600--800 km,
i.e., in the region where granulation structures are weakened but still visible on Doppler images. It can be
assumed that the altered structure of granulation in the lower chromosphere is one of the causes for the
reddening of shifts in the cores of strong lines. It is possible that the red contributions to the line from the
downward flows slightly prevail over the blue contributions from the upward flows, and the shift of the
core of the velocity-sensitive Mg line may be red.

\section{Line bisectors in stellar spectra}

{\bf Bisector noises and errors.} Individual bisectors in the spectra of the stars are distorted by noise to a greater
or lesser extent. Note that a slight tortuosity of the bisectors is visible even at very high resolution in the IAG
solar flux atlas (Fig.~1). The tortuous character is even more pronounced in the stars with $v\sin i \geq  3$  km/s. In
principle, radiation at any given point of the line profile comes from a wide range of depths in the photosphere
and total disk. This should lead to averaging and smoothing of all possible oscillations and to a
smoother shape of the bisectors. However, we see large observational noise in the line profiles and their
bisectors, which does not allow obtaining reliable results for individual bisectors. Small noise of rotationally
broadened profiles leads to even larger errors in the bisectors. In addition, if we proceed from the fact
that the tortuosity is more pronounced in stars with $v \sin i > 5$  km/s, it can be assumed that another cause
for this may be the redistribution of Doppler shifts caused by rotation in the surface regions near the limb,
as was noted in
\cite{1985PASP...97..543G}.  
In colder, slowly rotating stars, one can also see the chaotic appearance of the bisectors.
In addition to noise, there is uncertainty with blends, which also introduces additional scatter and
affects the reliability of the results. Averaging the bisectors makes it possible to remove noise; for this reason,
all the results of the analysis of stellar spectra were obtained based on the averaged bisectors.

 \begin{figure}[h!t]
 \centerline{
 \includegraphics   [scale=1.]{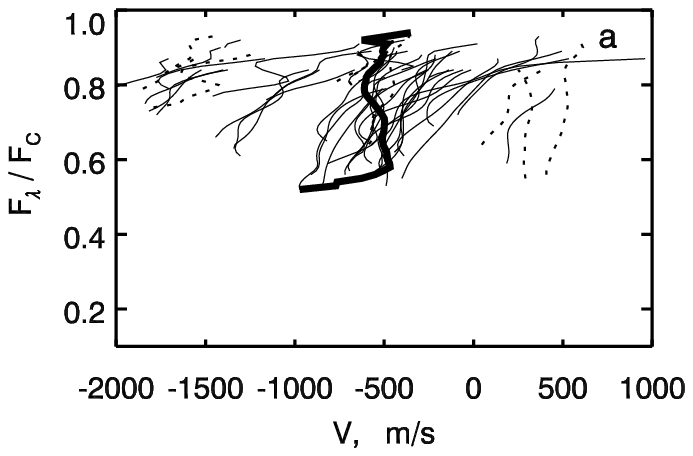}
 \includegraphics   [scale=1.]{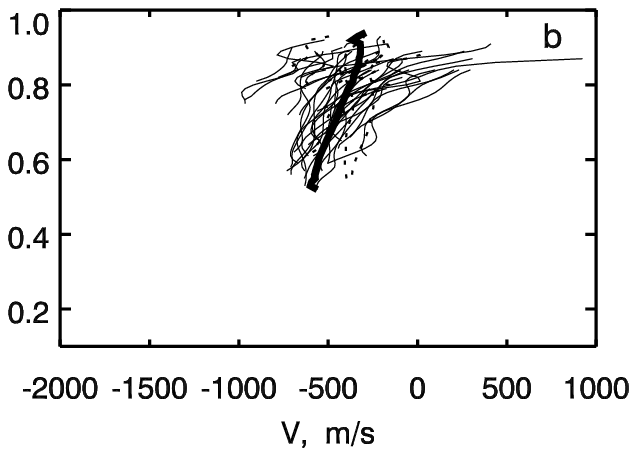}
     }
  \caption {\small
(a) Individual bisectors are shown on the Mg scale for the Fe I (thin solid lines)
and Fe II (dashed lines) lines and the
mean bisector (thick solid lines) for the star HD 102361; (b) the same bisectors
taking into account the correction for wavelength dispersion.
}
 \end{figure}
 \begin{figure}[h!]
 \centerline{
 \includegraphics   [scale=1.2]{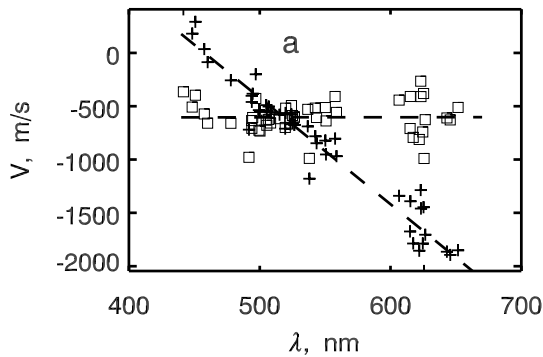}
     }
  \caption {\small
Convective line shifts are shown on the Mg scale in the spectrum of the star HD 102361 depending on the wavelength
(crosses) and shifts of the same lines taking into account the correction for wavelength dispersion (squares).
 }
 \end{figure}

For stars, we will consider all bisectors on the Mg scale. This means that their position has already been
corrected for the gravitational displacement and radial velocity of the star, and it corresponds to the convective
shift of the line core with an error equal to the convective shift of the Mg line in this star. The magnitude
of this convective shift is unknown, but it should be presumably small within the accuracy of
this analysis for solar-type stars.

 \begin{figure}[h!t]
 \includegraphics   [scale=1.01]{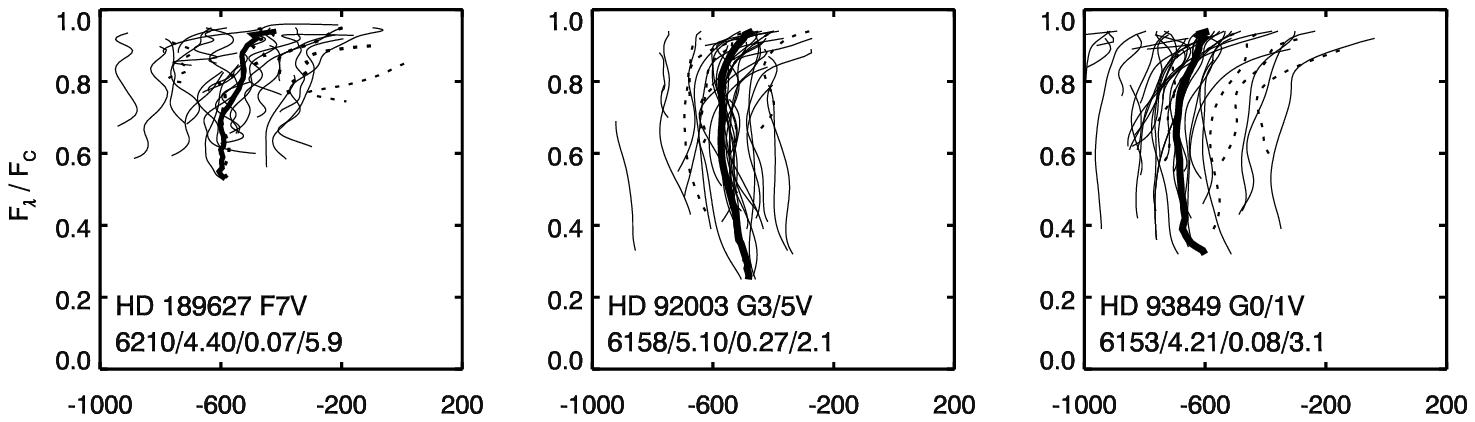}
 \includegraphics   [scale=1.01]{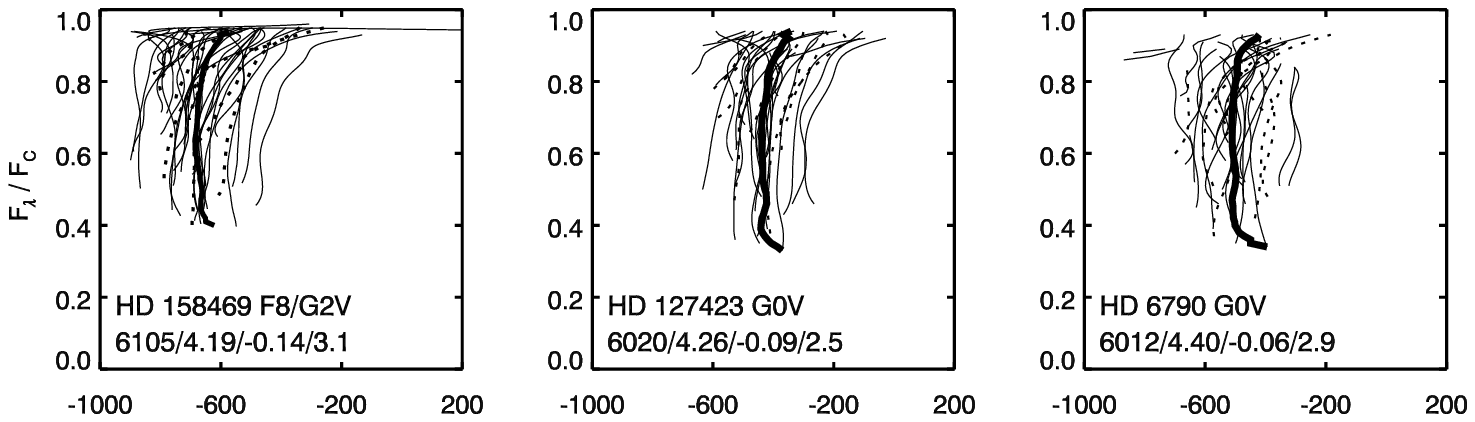}
 \includegraphics   [scale=1.01]{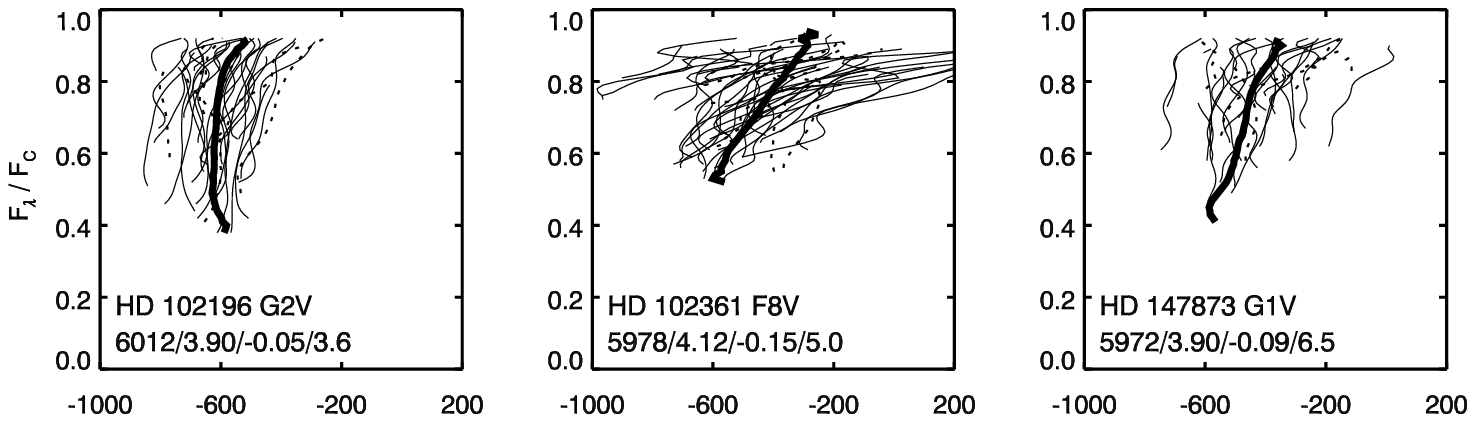}
  \includegraphics  [scale=1.01]{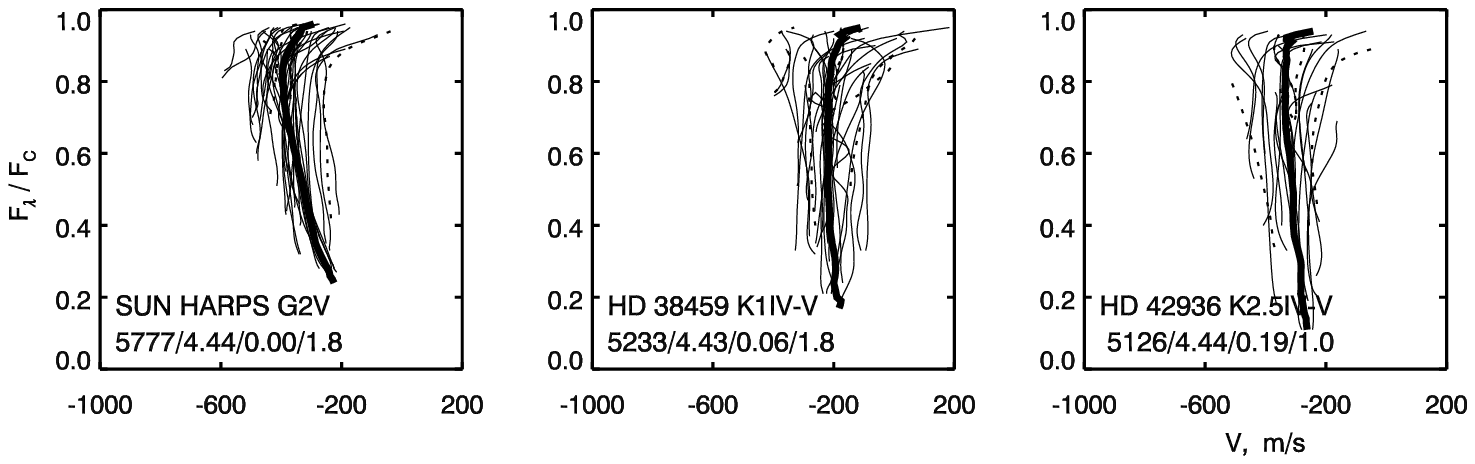}
  \includegraphics  [scale=1.01]{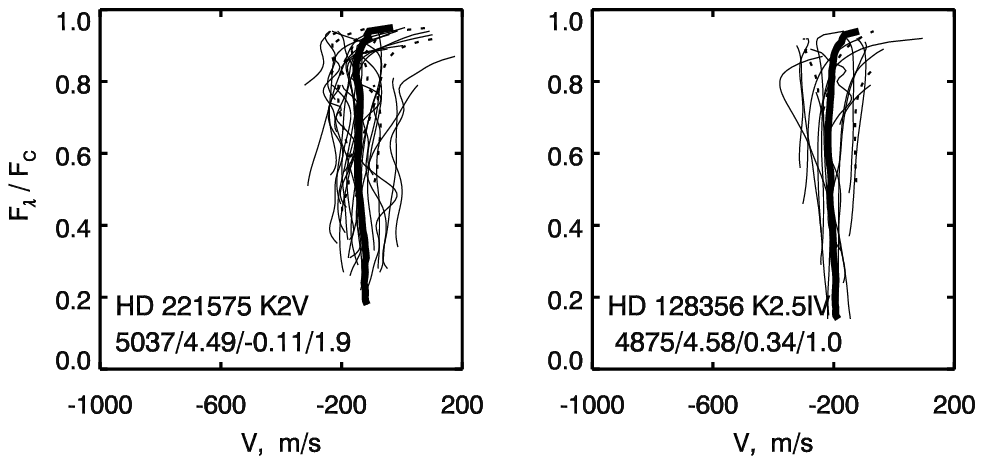}
  \caption {\small
Individual bisectors of the Fe I (thin solid lines) and Fe II (dashed lines)
lines as well as the averaged bisectors (bold
solid lines) in the Mg scale for solar-type stars with the indicated main
parameters $T_{\rm eff}{/}\log g/ {\rm[N/H]}{/}v \sin i$.
  }
 \end{figure}
 \begin{figure}[h!b]
 \centerline{
 \includegraphics   [scale=1.]{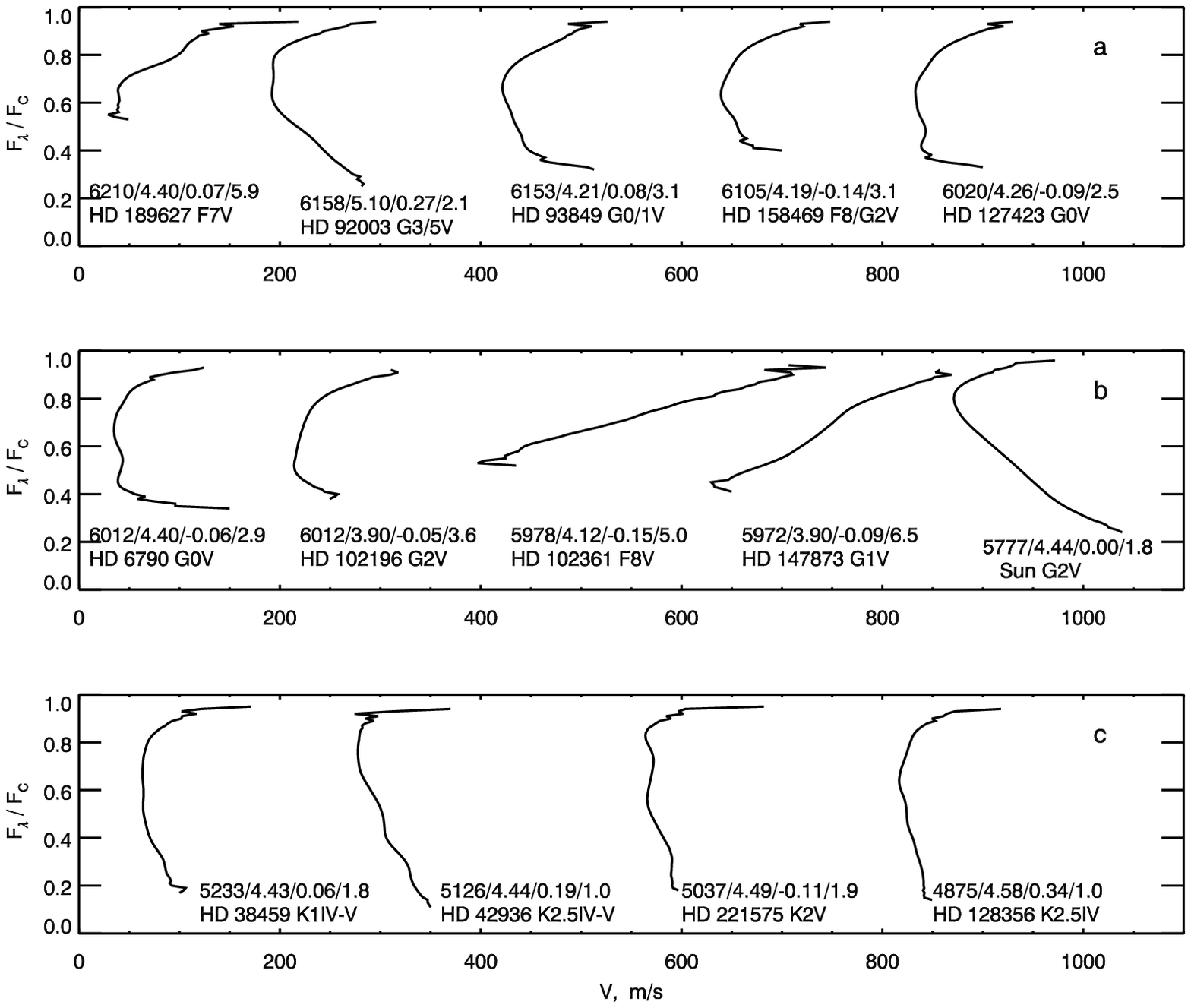}
     }
       \caption {\small
Changes in the shape of the mean bisectors with the effective temperature of the star. The horizontal arrangement of
the bisectors on the shift scale is arbitrary but ordered by the effective temperature. Star parameters $T_{\rm eff}{/}\log g/ {\rm[N/H]}{/}v \sin i$,
as well as the type of star, are noted under each bisector.
}
 \end{figure}

{\bf Line bisectors in the spectrum of the star HD 102361 F8V}. Our preliminary analysis of individual bisectors
showed a very wide range of bisector shifts in the spectrum of HD 102361, which is sharply different
from other stars. In addition, the magnitude of the shifts depends on the line wavelength (Fig. 6a). Blue
shifts reach approximately $-2000$  m/s in the 538--643 nm range, and red shifts reach almost 500 m/s in
the 441--460 nm range, i.e., the line shifts become bluer with increasing wavelength. This is clearly seen
in Fig. 7, which shows the dependence of the shifts of the individual lines on the wavelength. This dependence
resembles the dependence found in the HINKL atlas (see Fig. 3). In this regard, we introduced a dispersion
correction to the wavelength scale for this star. Figure 7 shows the corrected line shifts, and Fig. 6b
shows the positions of the bisectors of these lines in the corrected shift scale for the star HD 102361.

{\bf Changes in the bisectors with star temperature.} Figure 8 shows the bisectors of the lines of all the stars
and the Sun as a star (HARPS atlas) in order of decreasing effective temperature from left to right and from
top to bottom on the same scale. We see no significant differences in the bisectors of the Fe I and Fe II lines,
although it is known that the bisectors of strong Fe II lines for the Sun have a more pronounced C shape
than the bisectors of Fe I lines of the same strength. Apparently, this is a consequence of the lower resolution,
higher noise, and a small number of Fe II lines. If we consider the bisector of the strong line from
the top near the continuum to the bottom near the core, we can get an idea of the variation in the total
velocity of upward and downward flows from bottom to top in the photosphere of a particular star. Any
point on the bisector reflects the average velocity of convective flows at photospheric heights
from which radiation is emitted at the corresponding point of the line profile. To more accurately tie this velocity to
the height, it is necessary to calculate the profile of the line formation heights. However, even one bisector
can tell a lot.

Figure 8 shows that, in general, the individual bisectors of the weakest lines are most blue-shifted on
the shift scale, since these lines form in deep layers, where the convective velocities of ascending
granules are large. The strong lines are less displaced because their cores form in the high layers of the photosphere,
where granulation is significantly weaker. The placement of the bisector shifts on the scale from the weakest
to the strongest lines shows a certain range of convective velocities inherent in a particular star. The
width of this range increases with effective temperature and with decreasing surface gravity and metallicity.
We can say that this range of shifts is one of the signatures of the efficiency of convection in the stellar photosphere.

The mean bisectors, as seen in Fig. 9, have a C shape for most stars, same as in the case of the Sun,
but with different bulge in the middle of the bisector, the size of which depends on the main parameters
of the star. Among all the bisectors, the bisectors for the stars HD 189627 F7V, HD 102361 F8V, and
HD 147873 G1V with a rotation velocity $v\sin i \geq 5$  km/s stand out the most. The shape of these three
bisectors resembles a greatly elongated top of the letter C or even a slash symbol (/). This shape is caused
by the effects of rotation on the line profiles. As is known from
\cite{1985PASP...97..543G},  
the lower part of the bisector compresses
or contracts with an increase in the rotation velocity and deviates more and more towards the
blue part of the spectrum; for this reason, the shifts of the line cores in such stars become bluer.

Mean bisectors differ in length. In the spectra of hotter stars, absorption lines are significantly weakened
as compared to cooler stars; their bisectors become shorter and their cores more displaced since they
form in deeper layers. As a rule, the hotter the star, the faster it rotates. The length of the mean bisector
depends not only on temperature but also on the star's rotation velocity and metallicity. This is especially
noticeable for stars with large $ v \sin i$, e.g., the mean bisector in the star HD 189627 is much shorter than
the others and the shift of its lower part is bluer.
 \begin{figure}[h!b]
 \centerline{
 \includegraphics   [scale=1.]{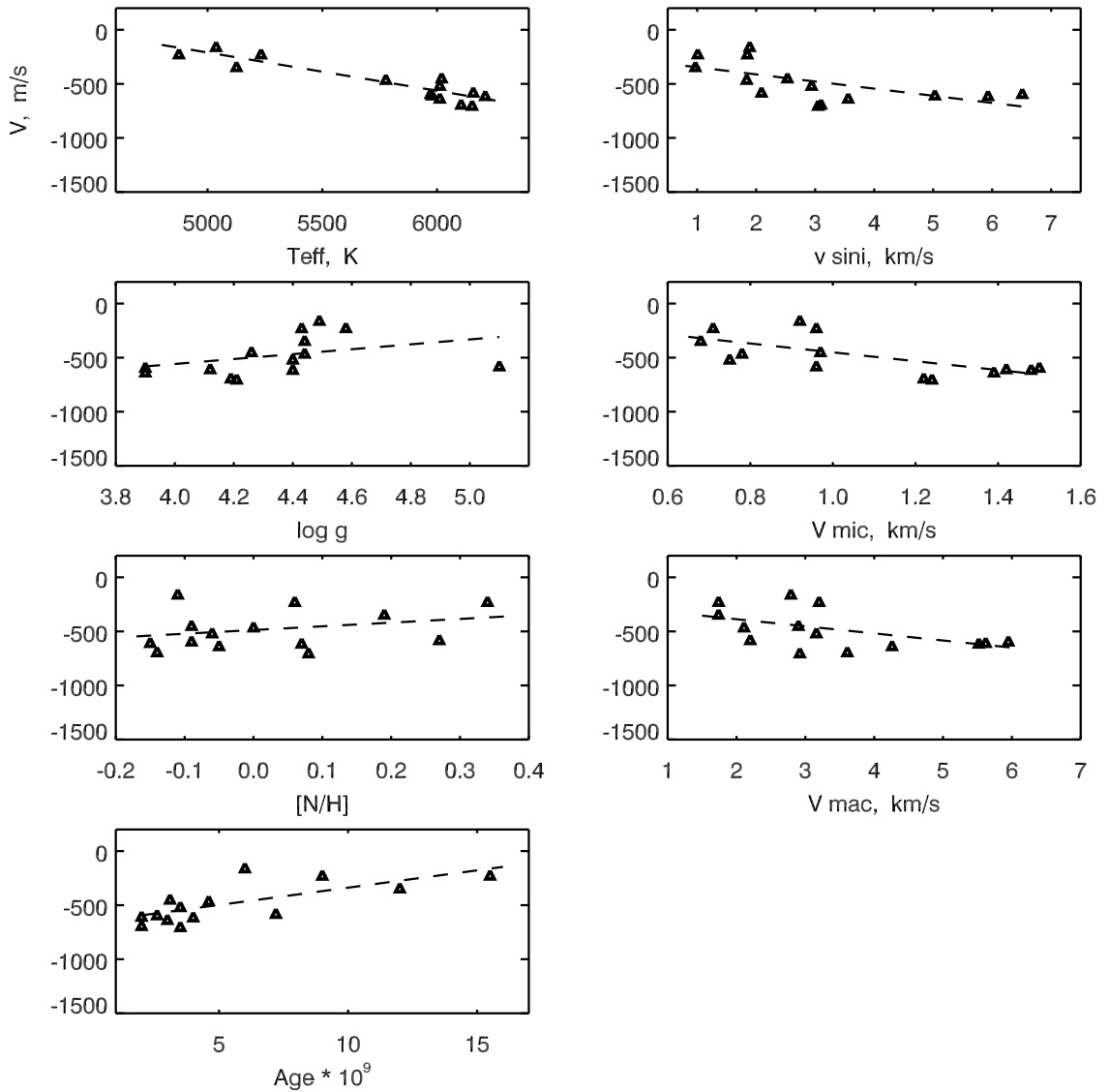}
 }
  \caption {\small
Average convective velocities derived from the bisector span depending on the effective temperature, gravity,
metallicity, age of the star, rotation velocity, and micro- and macroturbulence
velocities (the dashed lines show the linear approximation).}
 \end{figure}

An important characteristic of the bisector is its span or shift of the bluest point. It can be seen from
Fig. 9 that the span depends on the temperature. The higher the effective temperature,
the greater the curvature
 or span of the bisector and the greater the average convective velocity. In addition, the span depends
on the surface gravity, which tends to weaken the motion of the ascending granules and, due to this, the
span decreases in stars with higher gravity. In metal-rich stars, the opacity of the photosphere increases,
the motion of the ascending granules slows down, and the span becomes smaller. For example, in the stars
HD 93849 and HD 158469 with close $T_{\rm eff}$, $\log g$, and $v \sin i$  values, the mean bisectors
differ due to different
metallicities. In this regard, we cannot expect a clear dependence of the span of the bisectors on temperature
in our sample of stars, since the other parameters vary from star to star, albeit not much. In general,
we can say that the span of the bisector depends on the structure of the atmosphere, which is determined
by the main parameters of the star.

The height of the span of the mean bisector is also a characteristic feature. As can be seen from Fig. 9a,
the hotter the star, the smaller the span height. This means that effective granulation in hot stars extends into
the higher layers of the photosphere. In colder stars (Fig. 9c), the span height is large and this indicates that
the effective ascent of matter in the granules occurs in deeper layers than in the case of hot stars. Hence, it
follows that the higher the temperature, the higher the convective flows of hot matter rise into the atmosphere.

{\bf Convective velocities in stellar photospheres.} We measured the shift of the bluest point on the mean
bisector, i.e., the span of the bisector, and obtained the average convective velocity in the photosphere for
each star. The results are presented in Table 1 and Fig. 10 depending on the parameters of the star. It is
clearly seen that the closest correlation exists between the average convective velocity or the ascent velocity
of granules and the effective temperature of the star. The rest of the dependences shown in Fig. 10 have a
greater scatter of values; however, a certain tendency is observed everywhere in accordance with the physical
sense. We would like to note the existence of a correlation between the convective velocity and the
micro- and macroturbulence parameters. It is not without reason that the macroturbulence velocity is
often called the first signature of granulation in a star's atmosphere. The higher the ascent velocity of granules,
the more efficient the granulation and the stronger the micro- and macroturbulence in stars, i.e., the
stronger the dispersion of convective velocities. The influence of the star's rotation on the line bisectors is
also confirmed. The higher the rotation velocity, the higher the ascent velocity of the granules. The correlation
between the mean convective velocity and the age of the star should also be noted. It is obvious
that granulation efficiency should decrease with age. Thus, none of the dependences presented in Fig. 10
contradict the physical sense and they confirm the reliability of observational data on stellar asymmetry.

\section{Conclusions}

In this study, we have analyzed the asymmetry of lines in the spectra of solar-type stars obtained with the
HARPS spectrograph with a resolution of approximately 120000 and a signal-to-noise ratio above 100.
Bisectors, which are chains of convective blue shifts in the photospheric layers where the corresponding
points of the spectral line profile are formed, were measured for the selected lines. The main findings are as
follows.

In solar-type stars, the bisectors of very weak lines show blue shifts in the deepest layers of the photosphere,
and their shape resembles the upper part of the letter C. The bisectors of strong lines carry information
about the convective velocities of almost all the photospheric layers, and they have a shape similar
to the letter C for most stars, which becomes more like the / symbol if the rotation velocity of the star
$v \sin i \geq 5$ km/s.

With an increase in the effective temperature, the curvature of the bisector systematically increases.
This is confirmed by the clear dependence of the magnitude of the bisector span on the effective temperature.
The mean convective velocities derived from the magnitude of the bisector span increase from $-150$
to $-700$ m/s in stars with an effective temperature of 4800 to 6200 K, respectively. With increasing surface
gravity and metallicity, convective velocities become lower. As the star ages, convective velocities
decrease, as expected. The photospheric macroturbulent velocity, which is the dispersion of convective
velocities, directly depends on the magnitude of these velocities.

The line bisectors in the solar flux spectrum presented in the HARPS atlas fit well into the general picture
of the bisectors of solar-type stars. The comparative analysis of the bisectors measured in the absolute
scales of four solar flux atlases showed that the wavelength scale in the FTS atlases IAG, NECKL, and the
HARPS atlas is set with an accuracy of $\pm 10$ m/s in the range of 450--650 nm. The wavelength scale in the
HINKL atlas is shifted relative to the IAG atlas by an average of $-240$ m/s. In addition, the HINKL scale
shift changes with wavelength from $-100$ to $-330$ m/s in the range of 450--650 nm, respectively. In this
regard, it is undesirable to use the HINKL atlas to determine the convective blue line shifts.

The above conclusions confirm the known facts regarding the nature of the line bisectors in the spectra
of solar-type stars and show that, in general terms, the stellar line bisectors are similar to the solar line
bisectors but with their own characteristic features. When we compare the shape of the bisectors and the
magnitude of the blue shifts of the lines of different strengths in the stellar spectrum with solar data, it is
possible to obtain preliminary estimates of the average convective velocity, height of penetrating convection,
rotation velocity, and macroturbulent velocity in the stellar photosphere. By the magnitude of the
bisector span, it is possible to estimate the layers of the maximum effect of granulation on the profiles of
spectral lines and the energy of granulation in the photosphere.

Observation noise, blends, and insufficiently high spectral resolution distort the shape of the bisectors.
This affects the bisectors of rapidly rotating stars to a greater extent. Due to these reasons, the individual
bisectors are not reliable enough, so all the conclusions were drawn on the basis of the mean bisectors.
The mean convective velocities derived from the bisector span are only the lower limit since insufficiently
high resolution reduces this span. In addition, convective velocities in the scale of shifts relative to the Mg
line may contain the Mg line shift itself if it turns out to be nonzero. The values of the obtained average
convective velocities do not exceed the accuracy of laboratory wavelengths, which is 30--60 m/s at best.
To improve the analysis results, it is necessary to have more accurate laboratory wavelengths and stellar
spectra with higher resolution and signal-to-noise ratio.

{\bf ACKNOWLEDGMENTS}

I am sincerely grateful to Ya. Pavlenko and A. Ivanyuk for providing the observed spectra of stars and would also
like to thank the reviewer for important remarks.

\end{document}